\documentclass{emulateapj}

\def\arcdeg{\hbox{$^\circ$}}
\def\arcmin{\hbox{$^\prime$}}
\def\arcsec{\hbox{$^{\prime\prime}$}}
\def\deg2{\hbox{$\rm deg^{2}$}}

\def\fs{\hbox{$.\!\!^{\rm s}$}}

\def\farcs{\hbox{$.\!\!^{\prime\prime}$}}
\def\lsim{\mathrel{\rlap{\lower4pt\hbox{\hskip1pt$\sim$}}\raise1pt\hbox{$<$}}}                
\def\gsim{\mathrel{\rlap{\lower4pt\hbox{\hskip1pt$\sim$}}\raise1pt\hbox{$>$}}}                

\lefthead{Drake et~al.}
\righthead{Energetic Supernova 2008fz}

\begin{document}
\title{Discovery of the Extremely Energetic Supernova 2008fz}
\author{A.J. Drake\altaffilmark{1}, S.~G. Djorgovski\altaffilmark{1}, J.~L. Prieto\altaffilmark{2,3}, 
A.~Mahabal\altaffilmark{1}, D.~Balam\altaffilmark{4}, R.~Williams\altaffilmark{1},\\ M.~J. Graham\altaffilmark{1}, 
M.~Catelan\altaffilmark{5}, E.~Beshore\altaffilmark{6} and S.~Larson\altaffilmark{6}
}
\altaffiltext{1}{California Institute of Technology, 1200 E. California Blvd, CA 91225, USA}
\altaffiltext{2}{Department of Astronomy, Ohio State University}
\altaffiltext{3}{Carnegie Observatories, 813 Santa Barbara St., Pasadena, CA 91101}
\altaffiltext{4}{Dominion Astrophysical Observatory, National Research Council of Canada}
\altaffiltext{5}{Pontificia Universidad Cat\'olica de Chile, Departamento de Astronom\'ia y Astrof\'isica, 
Av. Vicu\~na Mackena 4860, 782-0436 Macul, Santiago, Chile}
\altaffiltext{6}{The University of Arizona, Department of Planetary Sciences,  Lunar and Planetary Laboratory, 1629 E. University Blvd, Tucson AZ 85721, USA}


\begin{abstract}
  
  We report on the discovery and initial observations of the energetic type IIn supernova (SN), 2008fz. The optical
  energy emitted by SN 2008fz (based on the light curve over a 88 day period), is possibly the most ever observed for a
  supernova ($1.4\!\times\!10^{51}$ erg).  The event was more luminous than the type IIn SN 2006gy, but exhibited same
  smooth, slowly evolving light curve.  As is characteristic of type IIn SN, the early spectra of 2008fz initially
  exhibited narrow Balmer lines which were replaced by a broader component at later times.  The spectra also show a blue
  continuum with no signs of Ca or Na absorption, suggesting that there is little extinction due to intragalatic dust in
  the host or circumstellar material.  No host galaxy is identified in prior coadded images reaching $R \sim 22$. From
  the supernova's redshift, $z=0.133$, we place an upper limit on the host of $M_R=-17$.  The presence of the SN within
  such a faint host follows the majority of recently discovered highly luminous SN. A possible reason for this
  occurrence is the very high star formation rate occurring in low-mass galaxies in combination with the low metallicity
  environment, which makes the production of very massive stars possible. We determine the peak absolute magnitude of
  the event to be $\rm M_V = -22.3$ from the initial photometry and the redshift distance, placing it among the most
  luminous supernovae discovered.

\end{abstract}

\section{Introduction}

The recent discovery of highly energetic type IIn supernovae such as SN 2006gy (Quimby 2006; Smith et al. 2007) and type
II-L SN 2005ap (Quimby et al. 2007) has stirred interest in the energy sources and fate of the most massive and luminous
stars.  The bulk of the energy expelled from a supernova is in the form of neutrinos (for a recent review see Haxton
2008). However, although experiments are underway to measure such neutrinos (Kowarik et al. 2009, Franckowiak et al.
2009), the amount of energy is not as readily measured for distant SN as is their optical emission.  The amount of
optical energy in the explosion of SN 2006gy was found to require a progenitor with mass $\rm >40M_{\sun}$ and suggests
either the explosion of a luminous blue variable (LBV) like $\eta$ Carinae, or the pair-instability destruction of a
very massive, low metallicity star (Woosley et al.~2007; Ofek et al.~2007).  Depending on their mass,
pair-instability SNe are expected to either directly form a black hole or completely disintegrate as they undergo
thermonuclear runaway (Heger \& Woosley 2002).  Such events are also of relevance for links between population III SNe
and long-timescale GRBs (Stanek et~al. 2003).

Recent direct evidence for the destruction of massive blue variable stars comes from an average luminosity
type IIn SN, 2005gl. This event was first noted to be due to a $\eta$ Carinae-like star by Gal-Yam et al. (2007).
Comparison of pre and post-explosion HST data shows the disappearance of progenitor with a luminosity
consistent with a very massive star (Gal-Yam \& Leonard 2009). Evidence for the pair-instability SN comes from the 
slow-evolving type Ic SN 2007bi discovered by the SNfactory (Nugent et~al. 2007). 
This event was found to have a core mass of $\rm\sim 100 M_{\sun}$, placing it well within 
the most massive star regime (Gal-Yam et al. 2009).
Other luminous type II SN, such as 2006tf, suggest there may be a transition between luminous type IIn events 
powered by interaction with CSM, to events where the opacity of the surrounding circumstellar material (CSM) 
confines the energy until it can diffuse out (Smith et al. 2008).

The brightest commonly observed population of core-collapse (CC) supernovae come from massive stars.  Extremely
luminous supernovae (hereafter ELSN) are more luminous than regular CC events which can range in peak
from $\rm M_{B} \sim -17$ to $\sim -20$ with peak magnitudes vary widely depending on their type 
(Richardson et al.~2002) 
For the purposes of this paper we define ELSN as supernova that were brighter the $\rm M_V = -21$ at peak. This groups
is made up of fewer than a dozen events discovered and confirmed in approximately as many years.  The recent ELSN have
all come from surveys covering large areas (rather than bright nearby galaxies), such as ROTSE (Quimby et al.  2007),
the SNfactory (Gal-Yam et al.  2009) and CRTS (Drake et al.  2009a). In most cases these have been discovered in
intrinsically faint galaxies.  This is interesting as such low-mass galaxies tend to have the low metallicity (Forbes,
S\'anchez-Bl\'azquez \& Proctor 2005; Spolaor et al.~2009) which may favor the formation of the massive star progenitors
to hypernovae (Heger et al.~2003).  Hypernovae exhibit broad-line type Ic SN spectra (Sahu et~al. 2009), and are
believed to be linked to the long-timescale gamma-ray bursts during the deaths of massive stars (Paczynski 1997).  Such
GRBs are found in low-mass, low-metalicity star-forming galaxies (Fruchter et al. 2006; Woosley \& Heger 2006).  Some
ELSN are found in the same environments as hypernovae, for example SN 2007bi was a bright broad-line type Ic in a faint
low metalicity galaxy (Young et al. 2009). However, as with 2007bi, not all broad-line type Ic events have been
associated with GRBs, nor do they all exhibit extremely luminosities.  Additionally, low-mass galaxies appear to have
much higher SN rates than expected for their baryonic content because of elevated star formation rate (Mannucci et al.
2005; Sullivan et al.  2006; Zheng et al 2007), although it is not known whether this extends to very faint galaxies.

\section{Data Reduction and Discovery}

The Catalina Real-Time Transient Survey (CRTS; Drake et al. 2009a) is a synoptic optical survey which searches for
transients in data taken by the Catalina Sky Survey (CSS; Larson et al. 2003). Results from the first six months of the
CRTS survey appear in a recent paper by Drake et al. (2009a) which includes transients found in a repeatedly imaged
region of $\sim 20,\!000 \rm deg^{2}$. 
The initial survey is based solely on CSS Schmidt telescope observations, which cover 1200 squares degrees of the sky per
night, four times, to a V magnitude of 20 with $2.5\arcsec$ resolution.  All data are automatically processed in
real-time, and OT discoveries are immediately distributed publicly as VOEvents, HTML and 
RSS\footnote{VOEventNet, http://voeventnet.caltech.edu/ and SkyAlert, http://www.skyalert.org/}.

As of July 2009 over a 850 unique optical transients\footnote{http://nesssi.cacr.caltech.edu/catalina/Allns.html}(OTs) have 
been discovered by searching for large variations between image epochs. These OTs include over hundred new cataclysmic
variables (CVs), as well as dozens of Blazars, flare stars and high-proper motion (HPM) stars and more than 140 supernovae.
The CRTS survey covers the sky irrespective of potential transient targets, unlike most current surveys 
for nearby supernovae which target between a few hundred and a few thousand bright nearby galaxies (KAIT/LOSS, Pucket/POSS, CHASE).
However, as transient detection is performed using catalogs, CRTS is most efficient at discovering supernovae that are 
more luminous than their hosts. This biases detection toward the brightest SN, such as ELSN (e.g., SN 2006gy; Smith et
al. 2007), and SNe in very faint galaxies (e.g., SN2009aq; Mahabal et al. 2009).

SN 2008fz was discovered by CRTS at $\rm \alpha = 23^{h} 16^{m} 16\fs57$ $(\pm0\fs02)$ 
$\rm \delta = +11 \arcdeg 42 \arcmin 47 \farcs 4$ $(\pm0\farcs2)$ (J2000) 
on September 22nd 2008 (Drake et al. 2008a). As with all CRTS transients, the object was rapidly classified based on
existing data and placed on public webpages. Photometric follow-up was taken with the Palomar 1.5m on September 23rd.
As there was no visible host galaxy the object was first classified as either a CV or SN (Drake et al. 2008b).  A noisy
spectrum was obtained September 23rd with the 1.82m Plaskett Telescope (range 390-703 nm, resolution 0.3 nm) and the
transient was initially classified as a type Ic SN (Hsiao et~al. 2008).  We obtained second spectrum with the SMARTS
1.5m on September 27th. However, this also had a low signal-to-noise ratio and exhibited no features strong enough to
classify it. The SN was independently rediscovered as a transient by the Palomar-Quest survey on September 29th and a
third spectrum was obtained on October 1st with the Palomar 5m telescope and DBSP (Mahabal et al. 2008).  Benetti et al.
(2008) subsequently took a spectrogram on October 9th using the TNG 3.5m telescope and DOLORES (range 350-930 nm,
resolution 1.3 nm), and reclassified the object as a type IIn SN at redshift $z=0.133$ based on narrow component of the
Balmer lines, and obtained a synthetic luminosity of $\rm M_R=-22.1$ from the spectrum.

\section{Photometry}

The CSS observations of SN 2008fz were taken unfiltered. V-magnitudes are routinely derived from CSS photometry by
calibrating each frame with between 10 and 100 G-dwarf stars in each field selected from 2MASS near-IR
data\footnote{http://www.ipac.caltech.edu/2mass/releases/allsky/doc/explsup.html}.  The magnitudes are transformed to V
following Bessell \& Brett (1988). Scatter in calibrated stars is typically $< 0.05$ magnitudes (Larson et al. 2007).

We checked the calibration for SN 2008fz using the standard star GD 246. This star lies within a degree of 2008fz, has a V
magnitude of $13.09 \pm 0.01$ (Landolt 2009). GD 246 was observed 16 times over four nights within minutes of observing
of SN 2008fz. Being a hot star this white dwarf it is expected to have a colour more similar to the SN than the G-dwarfs used
in routine calibration. The average CSS V-magnitude for GD 246 from the four overlapping nights is $13.07 \pm
0.04$. The observed magnitudes are clearly in excellent agreement.

In order to determine the absolute V-band magnitude of 2008fz the CSS photometry was corrected for Galactic extinction
of $A_V=0.136$, as derived from Schlegel et al (1998) reddening maps. However, no correction was applied for
intra-galaxy extinction within the host.  We apply K-corrections to account for the effective rest-frame bandwidth of
the redshifted supernovae (Oke \& Sandage 1968). However, as SN 2006gy and 2008fz photometry were taken unfiltered, we
do not apply corrections for the redshift of the SEDs.  Such corrections are expected to be small because of the
relatively flat flux distribution and low redshift of the SNe. 
The peak absolute magnitude is $V = -22.28\pm0.07$ for 
redshift ($z=0.133\pm0.003$), assuming $H_0 = 72$ km s$^{-1}$ Mpc$^{-1}$ and WMAP cosmology. 
Additional uncertainty is expected because of the unknown amount of intra-galaxy extinction.

\section{Spectra}

\begin{figure}[ht]{
\epsscale{0.8}
\plotone{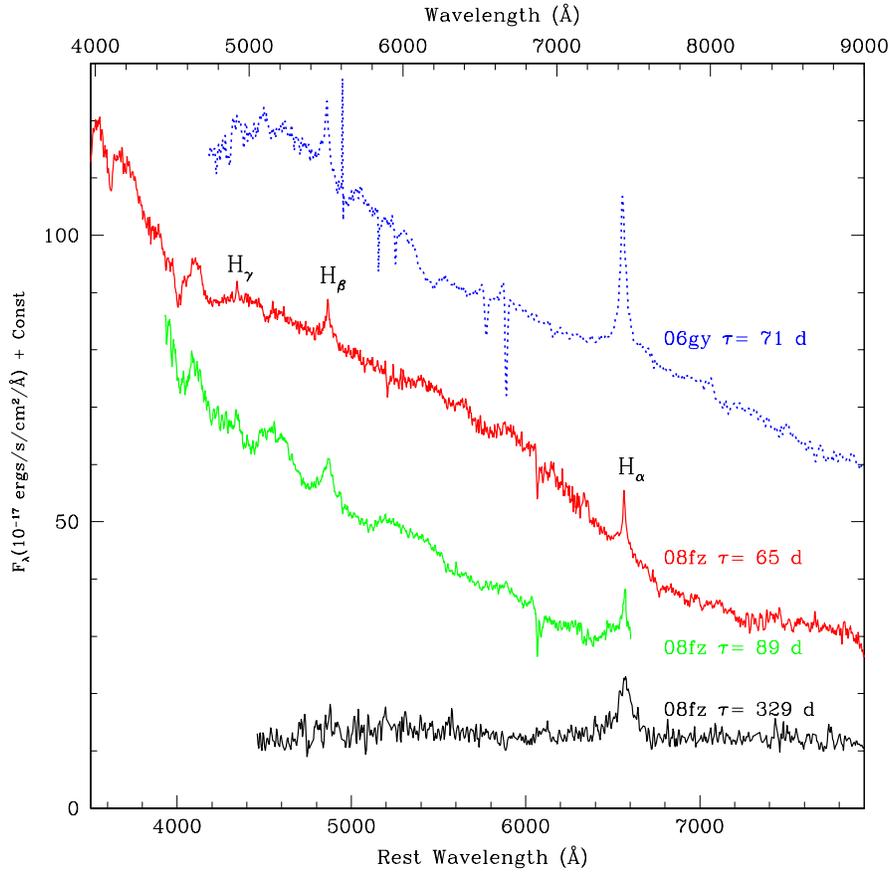}
\caption{\label{Spec}
Spectra of SN 2008fz at $t=65$d, $t=89$d and $t=329$d after explosion, obtained with the Palomar 5m + DBSP
MDM 2.4m +Modspec and Palomar 5m + DBSP, respectively. For comparison the $t=71$d spectrum of SN 2006gy 
from Smith et al. (2009) is shown with a dotted line. The explosion date for 2008fz assumes the event
has the same rise time as 2006gy. 
}
}
\end{figure}

As noted above, a spectrum of 2008fz was obtained with the Palomar 5m on October 1st UT. Additional spectra were obtained
on October 25th with the Modular Spectrograph (modspec) on the MDM 2.4m, and a later spectrum was obtained on June 22nd
2009 UT with the Palomar 5m telescope and DBSP (Oke \& Gunn 1982).  In Figure \ref{Spec}, we show spectra of SN 2008fz
along with that of SN 2006gy at a similar age.  The approximate time of 2008fz's explosion was determined assuming the
same rise time to maximum light as SN 2006gy. Therefore, this date uncertain by 10 days or more. A strong narrow
component to the Balmer lines are observed in SN 2008fz at early times, firmly placing the redshift at
$z=0.133\pm0.003$. At later times the continuum has become flat with broad $H_{\alpha}$ emission remaining as the only
strong feature in the 4500 - 8000 $\rm \AA$ range. The evolution of broad Balmer lines from initial narrow lines in 
type IIn supernovae is expected to be the sign of interaction with the CSM (Smith et al. 2009). 

\section{The Host Galaxy}

\begin{figure}{
\epsscale{0.8}
\plottwo{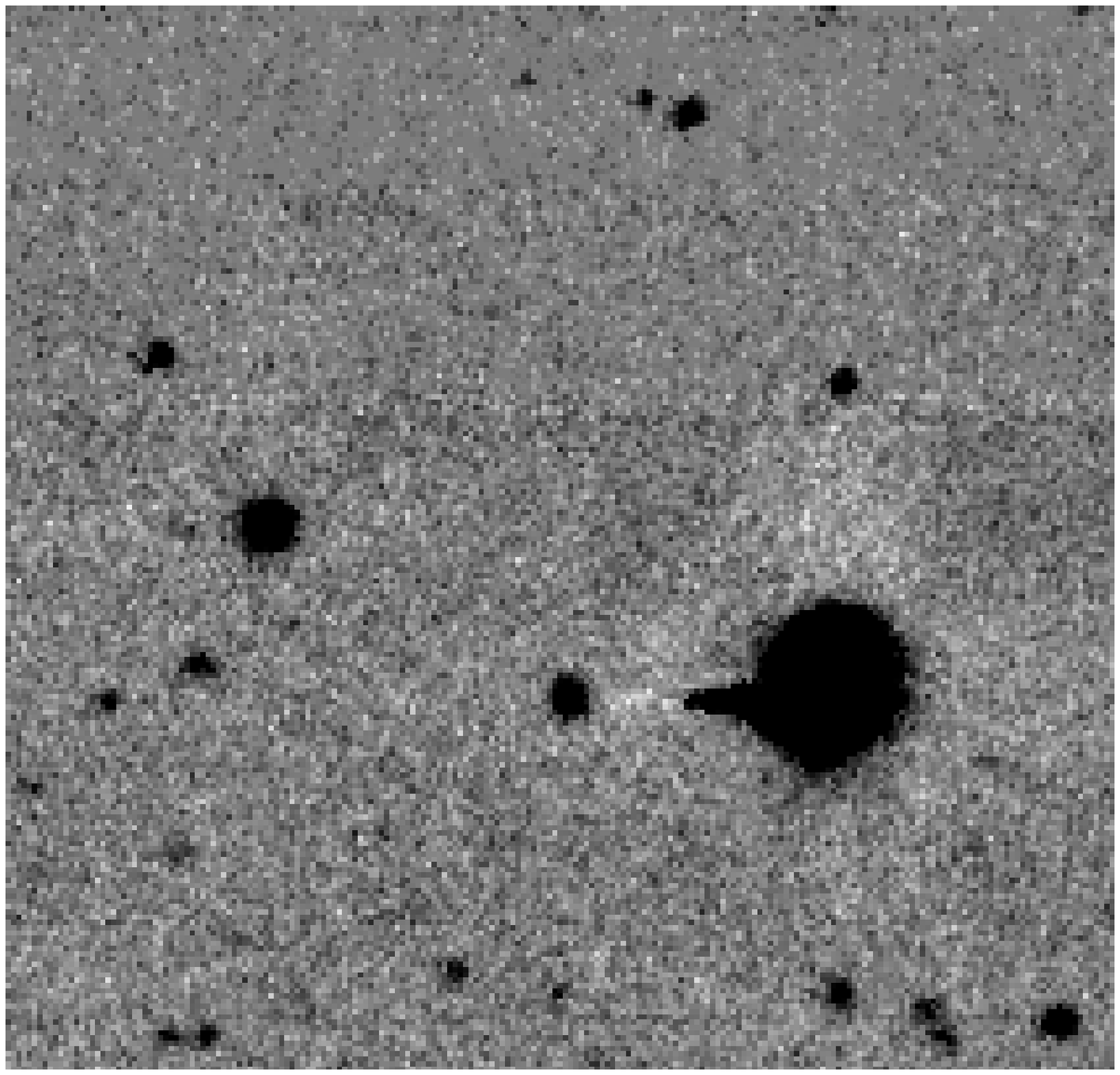}{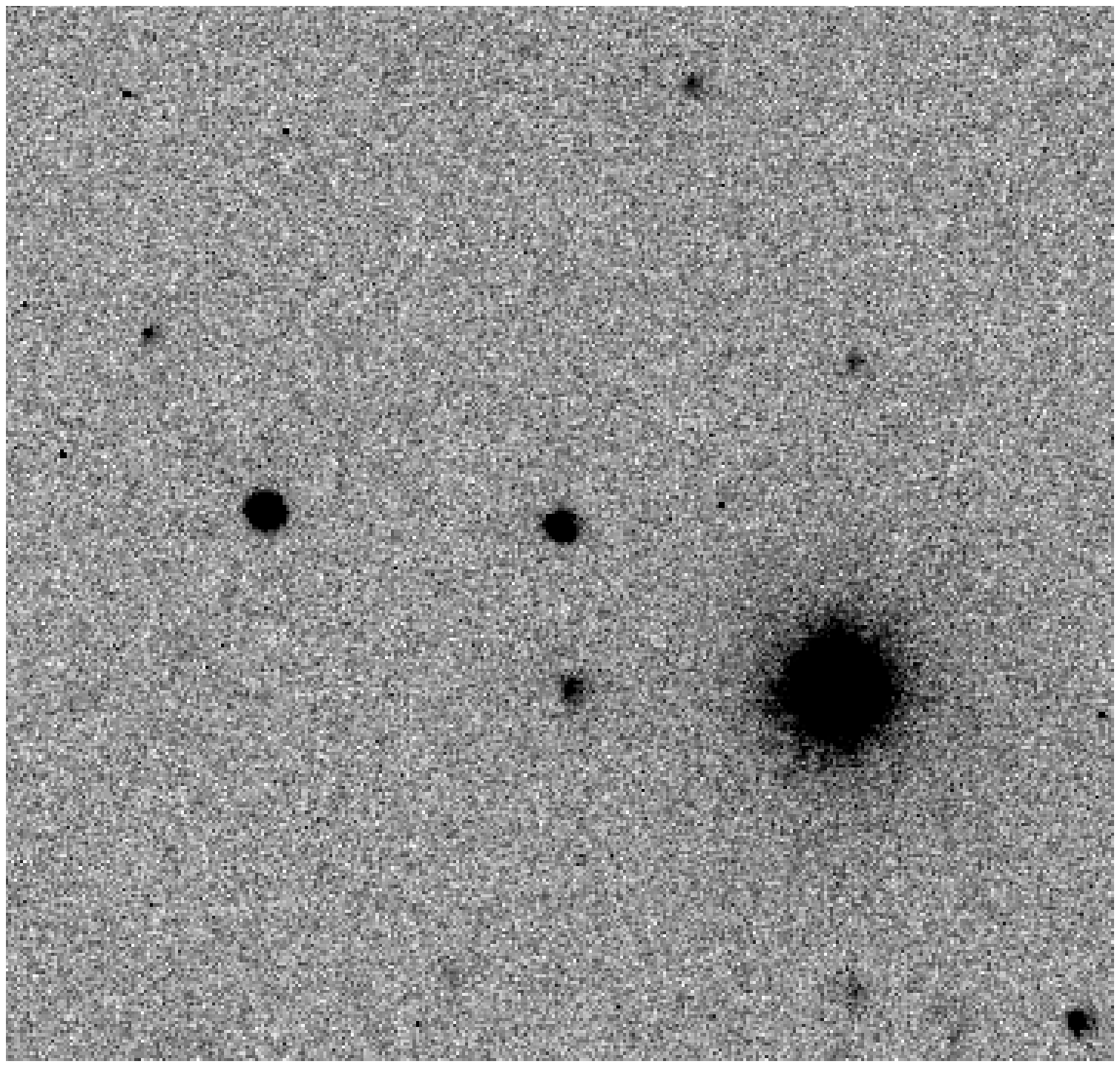}
\caption{\label{Img}
The Location of 2008fz.
Left: Palomar Quest Survey Johnson I-filter coadd image at location of the supernova.
Right: Palomar 1.52m Gunn-r follow-up image from September 23rd 2008 centered on SN 2008fz.
}
}
\end{figure}

In Figure \ref{Img} we present images showing a Palomar-Quest (PQ) coadd at a location of the SN and a September 23rd,
Gunn-r, Palomar 1.5m observation of the SN. The PQ coadd is constructed from nine, 140-second, Johnson I-band exposures
of the 1.2m Oschin Schmidt telescope taken on eight nights between August 31 2003 and September 24th 2006. The limiting 
magnitude is approximately 22.5. An R-band coadd image with a limiting magnitude of 22 also shows no sign of a host galaxy. 
Based on the SN redshift, the unseen host galaxy is fainter than $\rm M_{r} \sim -17$.

A single DSS-2 image (DSS2-B) of this region appears to show a small elongated object at the location of SN 2008fz. This
was suggested as a possible host (Drake et al. 2008a). However, close inspection of the image reveals an adjacent
negative patch of similar size and shape, suggesting that this is simply a photographic artifact. PQ B-band coadds of
similar depth show no such object.

\section{Comparison with SN 2006gy}

As noted earlier a number of highly luminous SNe have recently been discovered. These include highly luminous rapid
declining events such as SN 2005ap (Quimby et al. 2007) and 2008es (Gezari et al. 2009), as well as energetic
long-timescale events, such as SN 2006tf (Smith et al. 2008), SN 2007bi (Gal-Yam et al. 2009) and SN 2006gy (Smith et al.
2007).  Among these discoveries, the most optically energetic known event is SN 2006gy (Miller et al.  2009; Smith et al.
2009).

\begin{figure}{
\epsscale{0.6}
\plotone{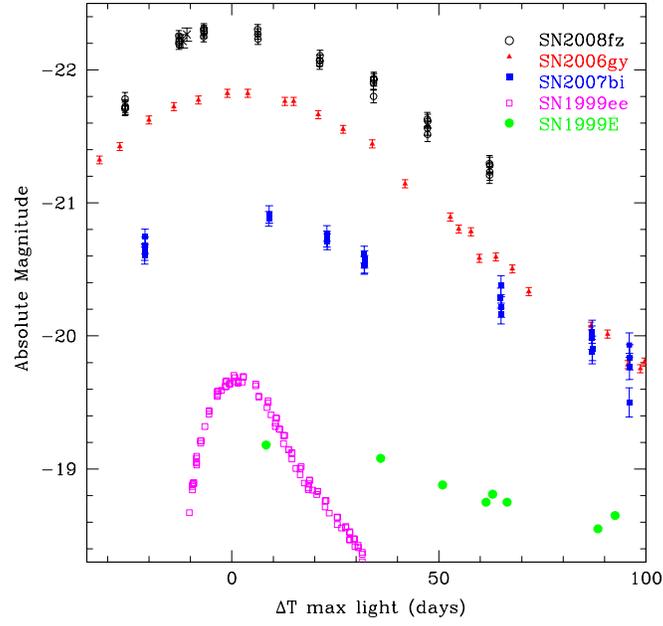}
\caption{
\label{LC2} 
Comparison of the absolute V magnitudes of SN 2008fz with ELSN and regular SN. 
Open circles, CSS magnitudes for 2008fz. Crosses, measurements of 2008fz 
made with the Palomar 1.5m. Triangles, KAIT pseudo-R magnitudes for SN 2006gy 
(Smith et~al. 2007) transformed to V. Filled boxes, CSS V-band magnitudes for ELSN 
SN 2007bi (Gal-Yam et~al. 2009). Open boxes, V-band magnitudes for the type Ia supernova 
1999ee (Stritzinger et~al. 2002). Large dots, V-band magnitudes for the type IIn supernova 
1999E (Rigon et~al. 2003). 
}
}
\end{figure}

In order to compare 2008fz with this event we transform the KAIT R-band proxy magnitudes given by Smith et al. (2007) to
V. The KAIT photometry are corrected for the very high amount of extinction from SN 2006gy's host galaxy (Smith et al.
2007), $A_R =1.25$, and the line-of-sight Galactic extinction, $A_R = 0.43$.  Such high extinction values are justified
by the presence of a red continuum relative to typical type IIn supernovae as well as deep Na {\scriptsize I} D
absorption lines. Smith et al (2007) assumed a standard extinction law when they derived the reddening for SN 2006gy.
Under this same assumption, the V-band Galactic extinction from Schlegel et al (1998) gives a total V-band extinction of 2.1 
magnitudes. As 2006gy is an unusual SN, and some of the extinction could be due to the CSM, it is not clear how accurate
this assumption is.  At peak magnitude Agnoletto et al. (2009) give the ($V-R$) colour of SN 2006gy as $0.57 \pm 0.14$.
We thus shift the KAIT values by the measured colour, while noting that, although the SN colour clearly evolves, both
KAIT and CSS data are taken clear and not corrected for colour evolution of the SED.  After transformation the KAIT
light curve V-band peak is $0.06$ magnitudes brighter than measured by Agnoletto et al. (2009).  This is well within
the $0.13$ mag uncertainty of Agnoletto et al (2009).  Early photometry data does not constrain the exact
outburst time of 2008fz, other than that the SN was not visible three months prior to detection to a limit of $V=20$.
Based on the similarity of the lightcurve to 2006gy we assume the rise time is similar to this event. Additionally, in
our analysis we only consider the photometric time span presented as the object became unobservable at later times on its
decline.  In Figure \ref{LC2} we present the light curves of SN 2008fz in relation to the ELSN
SN 2006gy (transformed KAIT photometry) and SN 2007bi (Gal Yam et al. 2009). 
For contrast we also display two supernovae with normal luminosity; the type IIn SN 1999E (Rigon et al. 2003),
and the type Ia SN 1999ee (Stritzinger et~al. 2002).
Integrating the light curve in Figure \ref{LC2} over the 88 days of photometric observations, correcting
for time dilation ($1+z$), and assuming zero bolometric correction, we calculate a total radiated energy of 
$E_{rad} = 1.4\pm0.2 \times 10^{51}$. Smith et al. (2007) found SN 2006gy radiated $1.2\pm 0.2 \times 10^{51}$ 
ergs for their full light curve. Over the same time window around maximum light we find that SN 2006gy radiated 
$1.0 \times 10^{51}$ ergs.

Over the first 65 days from maximum light SN 2006gy declined at an average rate of $\rm 0.020\pm0.001 mag.day^{-1}$
whereas 2008fz declined $\rm 0.018\pm0.001 mag.day^{-1}$ (corrected for the time dilation stretch). The combined optical
energy and decline rate results suggest that 2008fz was more energetic in its optical emission.  A more exact luminosity
and energy comparison between SN 2006gy and SN 2008fz is not possible without knowing the difference in wavelength
response of KAIT and CSS. However, 2008fz is $\sim 0.5$ magnitudes more luminous than SN 2006gy at peak.
As noted above, the peak luminosity of SN 2006gy is derived with a significant reddening correction.  Although there is
little sign of significant reddening from the host in the spectrum of SN 2008fz there may be additional extinction.  The
presence of such reddening would mean that absolute luminosity was underestimated, thus the gap in luminosity between 
2006gy and 2008fz would broaden.

\section{Results and Discussion}

The supernova 2008fz, discovered by CRTS, is possibly the most optically energetic supernova ever discovered.  Early and
late time spectra clearly show it is a type IIn event at $z =0.133$. Evidence suggests that this ELSN is significantly
brighter than the previous type IIn record holder, SN 2006gy (Smith et al. 2007).  However, SN 2008fz may not be the
most luminous supernova since SN 2005ap had an unfiltered peak absolute magnitude of $M=-22.7$ (Quimby et al. 2007).  
However, the V-band magnitude of Sn 2005ap would certainly be lower, and based on CSS data covering the supernova peak 
appears to have been fainter than SN 2008fz (Drake et al.~2009, in prep.).  An additional quickly declining supernova, 
SN 2008es, appears to have had a similar peak magnitude to SN 2008fz ($\rm M_V= -22.2$; Gezari et al. 2009, 
$\rm M_V= -22.3$; Miller et al 2008a).  Additional confirmed ELSN have also been discovered by CRTS including two
luminous type Ic supernova, 2008iu (Drake et al. 2009b) and the most luminous event, 2009de (Drake et al.  2009c).  
Both of these events are brighter than the type Ic proposed pair-instability supernova, 2007bi (Gal-Yam et~al.  2009).

Additional probable ELSN discovered by CRTS include CSS081009:002151-163204 (Drake et al. 2008c). This object was
spectroscopically observed by Silverman et al. (2008) and found to be a possible very luminous supernova or AGN at
$z=0.285$. Miller et al. (2008b) used SWIFT and PAIRITEL to rule out the AGN possibility. A 21st magnitude galaxy was
later discovered in the SDSS database at photometric redshift $z=0.20 \pm 0.09$. Based on CRTS and SWIFT photometry 
and a redshift of Silverman et al. (2008), the peak luminosity of this object was $\rm M_V= -22.3$.  Three other ELSN 
candidates have been discovered by CRTS but are yet to be spectroscopically confirmed.

The similarity of SN 2008fz to 2006gy suggests that there is a significant population of very bright, energetic
supernovae.  Prior imaging show that SN 2008fz must reside within a faint galaxy ($\rm M_V < -17$). This kind of
environment is expected for pair-instability supernovae with high-mass very low-metalicity progenitor stars (Woosley et
al. 2007). However, much more work is required to test whether SN 2008fz could be a pair-instability supernova. In the
case of SN 2006gy, the event occurred near the bulge or spheroid of the SO/a galaxy NGC 1260 (Bernardi et~al. 2002), 
and although Ofek et~al. (2007) noted that the galaxy is near solar metallicity, Smith et~al. (2007) note the likely presence
of ongoing star formation. It is also possible that SN 2006gy originated in an orbiting, low-metalicity dwarf galaxy
lying behind NGC 1260. This would explain the high level of extinction observed.  Alternately, both of these energetic 
SN could come from more massive versions of the luminous giant star that appears to be responsible for SN 2005gl
(Gal-Yam \& Leonard 2009).

The discovery of such optically energetic SNe in intrinsically faint, low-metalicity galaxies mirrors the discovery of
long-timescale GRBs associated with SNe, in low-mass star-forming galaxies (Stanek et~al. 2006; Savaglio et al.
2008). Under these simple considerations, these kinds of events could be linked. However, additional
work is required to quantify the role selection biases.

\acknowledgements

We would like to thank the anonymous referee for their comments which helped significantly improve this paper.
The CRTS survey is supported by the U.S. National Science Foundation under grants AST-0407448 and AST-0909182.  This
work is supported by the National Science Foundation under Grant No. CNS-0540369. The PQ survey is supported by the U.S.
National Science Foundation under Grants AST-0407448 and AST-0407297. The CSS survey is funded by the National
Aeronautics and Space Administration under Grant No. NNG05GF22G issued through the Science Mission Directorate
Near-Earth Objects Observations Program. J.L.P. is supported by NSF grant AST-0707982. Support for M.C. is provided 
by Proyecto Basal PFB-06/2007, by FONDAP Centro de Astrof\'{i}sica 15010003, and by a John Simon Guggenheim Memorial 
Foundation Fellowship.

\end{document}